\documentclass[pre,twocolumn]{revtex4}
\usepackage[cp850]{inputenc}
\usepackage{natbib}
\usepackage{amssymb,amsmath}
\usepackage{epsfig}
\usepackage{bm}
\usepackage{color}

\usepackage{graphicx}

\usepackage{color}

\newcommand\beq{\begin{equation}}
\newcommand\eeq{\end{equation}}
\newcommand\beqa{\begin{eqnarray}}
\newcommand\eeqa{\end{eqnarray}}
\newcommand{\dd}{\text{d}}

\newcommand{\al}{\alpha}

\begin{document}

\title{Influence of the first-order contributions to the partial temperatures on transport properties in polydisperse dense granular mixtures}

\author{Rub\'en G\'omez Gonz\'alez\footnote[1]{Electronic address: ruben@unex.es}}
\affiliation{Departamento de F\'{\i}sica,
Universidad de Extremadura, E-06006 Badajoz, Spain}
\author{Vicente Garz\'{o}\footnote[2]{Electronic address: vicenteg@unex.es;
URL: http://www.unex.es/eweb/fisteor/vicente/}}
\affiliation{Departamento de F\'{\i}sica and Instituto de Computaci\'on Cient\'{\i}fica Avanzada (ICCAEx), Universidad de Extremadura, E-06006 Badajoz, Spain}

\begin{abstract}

The Chapman--Enskog solution to the Enskog kinetic equation of polydisperse granular mixtures is revisited to determine the first-order contributions $\varpi_i$ to the partial temperatures. As expected, these quantities (which were neglected in previous attempts) are given in terms of the solution to a set of coupled integro-differential equations analogous to those for elastic collisions. The solubility condition for this set of equations is confirmed and the coefficients $\varpi_i$ are calculated by using the leading terms in a Sonine polynomial expansion. These coefficients are given as explicit functions of the sizes, masses, composition, density, and coefficients of restitution of the mixture. Within the context of small gradients, the results apply for arbitrary degree of inelasticity and are not restricted to specific values of the parameters of the mixture. In the case of elastic collisions, previous expressions of $\varpi_i$ for ordinary binary mixtures are recovered. Finally, the impact of the first-order coefficients $\varpi_i$ on the bulk viscosity $\eta_\text{b}$ and the first-order contribution $\zeta^{(1)}$ to the cooling rate is assessed. It is shown that the effect of $\varpi_i$ on $\eta_\text{b}$ and $\zeta^{(1)}$ is not negligible, specially for disparate mass ratios and strong inelasticity.

\end{abstract}

\draft
\date{\today}
\maketitle

\section{Introduction}
\label{sec1}

The understanding of transport processes occurring in polydisperse granular mixtures (namely, a mixture of smooth hard spheres with inelastic collisions) is still an exciting unsolved problem \cite{BP04,RN08,G19}. The reason for this challenging target is twofold: first, there is a large number of relevant parameters involved in the description of the granular mixtures; and second, there is a wide array of intricacies arising in the derivation of kinetic theory models. Thus, to gain some insight into the problem, the two most common simplifications employed in many of the pioneering papers on granular mixtures \cite{JM87,JM89,Z95,AW98,WA99} were (i) to consider mixtures constituted by nearly elastic particles and (ii) to assume the equipartition of the total granular kinetic energy in the homogeneous cooling state (namely, they assume that the zeroth-order contributions $T_i^{(0)}$ to the partial temperatures $T_i$ of each species are equal to the granular temperature $T$). On the other hand, the last assumption can be only justified for quasielastic collisions since the failure of energy equipartition in granular fluids \cite{GD99b,MP99} has been confirmed by computer simulations \cite{MG02,BT02,DHGD02,PMP02,KT03,WJM03,BRM05,SUKSS06} and observed in real experiments of agitated granular mixtures \cite{WP02,FM02}. The above papers have also shown that the departure of energy equipartition depends on the mechanical differences among the particles and the coefficients of restitution of the granular mixture.

The inclusion of energy nonequipartition effects on transport in granular mixtures has been considered in more recent papers of dilute \cite{GD02,SGNT06,GMD06,GM07,SNTG09,GMV13a} and moderate densities \cite{GDH07,GHD07,MGH12}. In particular, the results derived from the inelastic Enskog equation \cite{GDH07,GHD07,MGH12} cover some of the aspects not accounted for in previous studies. More specifically, (i) they are expected to be applicable for a wide range of coefficients of restitution (since they take into account the nonlinear dependence of the transport coefficients on the coefficients of restitution); (ii) they consider the impact of nonequipartition of granular energy on the Navier--Stokes transport coefficients; and (iii) they are valid for moderate densities. Thus, these works \cite{GDH07,GHD07,MGH12} subsume all previous studies for dilute \cite{GD02,SGNT06,GMD06,GM07,SNTG09,GMV13a} and dense quasielastic \cite{JM87,JM89,Z95,AW98,WA99} granular mixtures, which are recovered in the appropriate limits.

Nevertheless, the theory developed for dense gases \cite{GDH07,GHD07,MGH12} is based on a simplifying assumption. Although not explicitly stated, the results derived in Refs.\ \cite{GDH07,GHD07,MGH12} were obtained by neglecting the first-order
contributions $T_i^{(1)}$ to the partial temperatures $T_i$. The existence of a non-zero first-order contribution $T_i^{(1)}$ induces a breakdown of the energy equipartition, additional to the one appearing in the homogeneous cooling state (which is only due to the inelastic character of collisions). In fact, $T_i^{(1)}\neq 0$ in the case of ordinary dense mixtures (namely, a dense hard-sphere mixture with elastic collisions). Although the partial temperatures are not hydrodynamic quantities, their determination is interesting by itself. In addition, a careful analysis of the first-order contributions to the collisional part $\mathsf{P}^{\text{c}}$ of the pressure tensor and the cooling rate $\zeta$ (which accounts for the rate of kinetic energy dissipation due to inelastic collisions) shows that there are contributions to $\mathsf{P}^{\text{c}}$ and $\zeta$ coming from the coefficients $T_i^{(1)}$. Since the first-order contributions to the partial temperatures are proportional to the divergence of the flow velocity $\mathbf{U}$, then the coefficients $T_i^{(1)}$ are involved in the evaluation of both the bulk viscosity $\eta_\text{b}$ (proportionality coefficient between $\mathsf{P}^{\text{c}}$ and $\nabla \cdot \mathbf{U}$) and the first-order contribution $\zeta_U$ to the cooling rate (proportionality coefficient between $\zeta$ and $\nabla \cdot \mathbf{U}$).  The coupling between $T_i^{(1)}$ and $\eta_\text{b}$ was in fact already recognized in the pioneering papers \cite{KS79a,KS79b,LCK83} of the Enskog theory for multicomponent ordinary mixtures.


The question arises then as to whether, and if so to what extent, the conclusions drawn from Refs.\ \cite{GDH07,GHD07,MGH12} for $\eta_\text{b}$ and $\zeta$ may be altered when the above new ingredient (first-order contributions to the partial temperatures) is accounted for in the theory. In this paper we calculate $T_i^{(1)}$ and assess the impact of these coefficients on $\eta_\text{b}$ and $\zeta$ for granular mixtures.

The plan of the paper is as follows. The Enskog kinetic equation for polydisperse granular mixtures is introduced in Sec.\ \ref{sec2} and the corresponding balance equations for the densities of mass, momentum, and energy are recalled. Section \ref{sec3} deals with the evaluation of the first-order contributions to the partial temperatures. As expected, the coefficients $T_i^{(1)}$ are given in terms of the solution to a set of  linear integral equations. The leading term in a Sonine polynomial expansion is retained in Sec.\ \ref{sec4} to solve the above set and obtain the partial temperatures in terms of the parameter space of the problem. For the sake of illustration, a binary mixture is considered in Sec.\ \ref{sec5}. The results show that the impact of the coefficients $T_i^{(1)}$ on both the bulk viscosity and the cooling rate is not in general negligible and must be accounted for, specially for disparate mass ratios and strong dissipation. The paper is closed in Sec. \ref{sec6} with some concluding remarks.

\section{Enskog kinetic equation for polydisperse dense granular mixtures}
\label{sec2}

We consider an $s$-component granular mixture of inelastic hard disks ($d=2$) or spheres ($d=3$) of masses $m_i$ and diameters $\sigma_i$. The subscript $i$ labels one of the $s$ mechanically different components and $d$ is the dimension of the system. Spheres are assumed to be completely smooth so that, inelasticity of collisions is only characterized by the constant (positive) coefficients of restitution $\al_{ij}\leqslant 1$. The mixture is also assumed to be in the presence of the gravitational field and hence, each particle feels the action of the force $\mathbf{F}_i=m_i \mathbf{g}$, where $\mathbf{g}$ is the gravity acceleration. For moderate densities, the one-particle velocity distribution function $f_i(\mathbf{r}, \mathbf{v}, t)$ of component $i$ verifies the set of $s$-coupled nonlinear integro-differential Enskog equations:
\beq
\label{1}
\frac{\partial f_i}{\partial t}+\mathbf{v}\cdot \nabla f_i+\mathbf{g}\cdot \frac{\partial f_i}{\partial \mathbf{v}}=\sum_{j=1}^sJ_{ij}[\mathbf{r},\mathbf{v}|f_i,f_j],
\eeq
where the Enskog collision operator is \cite{G19}
\begin{widetext}
\beqa
\label{2}
J_{ij}\left[\mathbf{r}_1, \mathbf{v}_1|f_i,f_j\right]&=&\sigma_{ij}^{d-1}\int \dd\mathbf{v}_2\int \dd\widehat{\boldsymbol{\sigma}}\Theta\left(\widehat{\boldsymbol{\sigma}}\cdot\mathbf{g}_{12}\right)
\left(\widehat{\boldsymbol{\sigma}}\cdot\mathbf{g}_{12}\right)\Big[\alpha_{ij}^{-2}\chi_{ij}
(\mathbf{r}_1,\mathbf{r}_1-\boldsymbol{\sigma}_{ij})f_i(\mathbf{r}_1,\mathbf{v}_1'',t)f_j(\mathbf{r}_1-\boldsymbol{\sigma}_{ij},\mathbf{v}_2'',t)
\nonumber\\
& & -\chi_{ij}(\mathbf{r}_1,\mathbf{r}_1+\boldsymbol{\sigma}_{ij})f_i(\mathbf{r}_1,\mathbf{v}_1,t)f_j(\mathbf{r}_1
+\boldsymbol{\sigma}_{ij},\mathbf{v}_2,t)\Big].
\eeqa
\end{widetext}
In Eq.\ \eqref{1}, $\boldsymbol{\sigma}_{ij}=\sigma_{ij} \widehat{\boldsymbol{\sigma}}$, $\sigma_{ij}=(\sigma_i+\sigma_j)/2$, $\widehat{\boldsymbol{\sigma}}$ is a unit vector directed along the line of centers from the sphere of component $i$ to that of component $j$ at contact, $\Theta$ is the Heaviside step function, and $\mathbf{g}_{12}=\mathbf{v}_1-\mathbf{v}_2$ is the relative velocity of the colliding pair. Moreover, $\chi_{ij}(\mathbf{r}_1,\mathbf{r}_1+\boldsymbol{\sigma}_{ij})$ is the equilibrium pair correlation function of two hard spheres, one of component $i$ and the other of component $j$ at contact, i.e., when the distance between their centers is $\sigma_{ij}$. The precollisional velocities $(\mathbf{v}_1'',\mathbf{v}_2'')$ are given by
\begin{equation}
\label{3}
\mathbf{v}_1''=\mathbf{v}_1-\mu_{ji}\left(1+\alpha_{ij}^{-1}\right)\left(\boldsymbol{\widehat{\sigma}}
\cdot\mathbf{g}_{12}\right)\boldsymbol{\widehat{\sigma}},
\eeq
\beq
\label{4}
\mathbf{v}_2''=\mathbf{v}_2+\mu_{ij}\left(1+\alpha_{ij}^{-1}\right)\left(\boldsymbol{\widehat{\sigma}}
\cdot\mathbf{g}_{12}\right)\boldsymbol{\widehat{\sigma}},
\end{equation}
where $\mu_{ij}=m_i/(m_i+m_j)$.

The first few velocity moments of the distributions $f_i$ define the hydrodynamic fields of the mixture. Thus, the local number density of component $i$ is
\beq
\label{4.1}
n_i=\int \dd \mathbf{v}\; f_i(\mathbf{v}),
\eeq
while the local mean flow velocity of grains is defined as
\beq
\label{4.2}
\mathbf{U}=\rho^{-1}\sum_{i=1}^s \int \dd \mathbf{v}\; m_i \mathbf{v} f_i(\mathbf{v}),
\eeq
where $\rho=\sum_i m_i n_i$ is the total mass density. Apart from the partial densities $n_i$ and the flow velocity $\mathbf{U}$, the other important hydrodynamic field is the granular temperature $T$. It is defined as
\beq
\label{4.3}
T=\frac{1}{n}\sum_{i=1}^s\int \dd\mathbf{v}\frac{m_{i}}{d}V^{2}f_{i}(\mathbf{v}),
\eeq
where $n=\sum_i n_{i}$ is the total number density and $\mathbf{V}=\mathbf{v}-\mathbf{U}$ is the peculiar velocity. At a kinetic level, it is also convenient to introduce the partial kinetic temperatures $T_i$ for each component. These quantities measure the mean kinetic energy of each component. They are defined as
\begin{equation}
\label{4.4}
T_i=\frac{m_{i}}{d n_i}\int\; \dd\mathbf{v}\;V^{2}f_{i}(\mathbf{ v}).
\end{equation}
According to Eq.\ \eqref{4.3}, the granular temperature $T$ of the mixture can be also written as
\beq
\label{4.5}
T=\sum_{i=1}^s\, x_i T_i,
\eeq
where $x_i=n_i/n$ is the mole fraction of component $i$.

An important property of the integrals involving the Enskog collision operator $J_{ij}[\mathbf{r},\mathbf{v}|f_i,f_j]$ is \cite{G19,GDH07}
\begin{widetext}
\beqa
\label{4.6}
I_{\psi_i}&\equiv & \sum_{i,j=1}^s\int\; \dd \mathbf{v}_1\; \psi_i(\mathbf{v}_1) J_{ij}[\mathbf{r}_1,\mathbf{v}_1|f_i,f_j]\nonumber\\
&=&\frac{1}{2}\sum_{i,j=1}^s\sigma_{ij}^{d-1}\int \dd\mathbf{v}_1\int\ \dd{\bf v}_{2}\int \dd\widehat{\boldsymbol{\sigma}}\,
\Theta (\widehat{{\boldsymbol {\sigma }}}\cdot \mathbf{g}_{12})
(\widehat{\boldsymbol {\sigma }}\cdot \mathbf{g}_{12})\Bigg\{\Big[\psi_i(\mathbf{v}_1')
+\psi_j(\mathbf{v}_2')-\psi_i(\mathbf{v}_1)-\psi_j(\mathbf{v}_2)\Big]
f_{ij}\left(\mathbf{r}_1,\mathbf{v}_1,\mathbf{r}_2,\mathbf{v}_2;t\right)\nonumber\\
& & +\frac{\partial}{\partial \mathbf{r}_1}\cdot \boldsymbol{\sigma}_{ij}\Big[\psi_i(\mathbf{v}_1')-\psi_i(\mathbf{v}_1)\Big]\int_0^1\; dx\; f_{ij}\Big(\mathbf{r}_1-x\boldsymbol{\sigma}_{ij},\mathbf{v}_1,\mathbf{r}_1+(1-x)\boldsymbol{\sigma}_{ij},\mathbf{v}_2;t\Big)\Bigg\},
\nonumber\\
\eeqa
\end{widetext}
where
\beq
\label{4.7}
f_{ij}\left(\mathbf{r}_1,\mathbf{v}_1,\mathbf{r}_2,\mathbf{v}_2;t\right)\equiv
\chi_{ij}(\mathbf{r}_1,\mathbf{r}_2) f_i(\mathbf{r}_1,\mathbf{v}_1,t)
f_j(\mathbf{r}_2,\mathbf{v}_2,t),
\eeq
$\psi_i(\mathbf{v}_1)$ is an arbitrary function of $\mathbf{v}_1$ and
\beq
\label{4.8}
\mathbf{v}_1'=\mathbf{v}_1-\mu_{ji}\left(1+\alpha_{ij}\right)\left(\boldsymbol{\widehat{\sigma}}
\cdot\mathbf{g}_{12}\right)\boldsymbol{\widehat{\sigma}}.
\eeq
The first term on the right hand side of Eq.\ \eqref{4.6} represents a collisional effect due to scattering with a change in velocities. The second term provides a pure collisional effect due to the spatial difference of the colliding pair. For elastic collisions, the first term vanishes. The balance equations for the densities of mass, momentum, and energy can be derived by using the property \eqref{4.6}. They are given by \cite{G19}
\begin{equation}
\label{4.9}
D_t n_i+n_i\nabla\cdot \mathbf{U}+\frac{\nabla\cdot\mathbf{j}_i}{m_i}=0,
\end{equation}
\begin{equation}
\label{4.10}
D_t\mathbf{U}+\rho^{-1}\nabla\cdot\mathsf{P}=\mathbf{g},
\end{equation}
\beq
\label{4.11}
D_tT-\frac{T}{n}\sum_{i=1}^s\frac{\nabla\cdot\mathbf{j}_i}{m_i}+\frac{2}{dn}
\left(\nabla\cdot\mathbf{q}+\mathsf{P}:\nabla\mathbf{U}\right)=-\zeta T.
\eeq
In the above equations, $D_t=\partial_t+\mathbf{U}\cdot\nabla$ is the  material derivative, $\rho_i=m_in_i$ is the mass density of component $i$,  and
\beq
\label{4.12}
\mathbf{j}_i=m_i\int\;\dd\mathbf{v}\; \mathbf{V}f_i(\mathbf{v})
\eeq
is the mass flux for component $i$ relative to the local flow $\mathbf{U}$. A consequence of the definition \eqref{4.12} of the fluxes $\mathbf{j}_i$ is that
\beq
\label{4.12.1}
\sum_{i=1}^s\; \mathbf{j}_i=\mathbf{0},
\eeq
and hence, only $s-1$ mass fluxes are independent. The pressure tensor $\mathsf{P}(\mathbf{r},t)$ and the heat flux $\mathbf{q}(\mathbf{r},t)$ have both kinetic and collisional transfer contributions, i.e.,
\beq
\label{4.13}
\mathsf{P}=\mathsf{P}^\text{k}+\mathsf{P}^\text{c}, \quad \mathbf{q}=\mathbf{q}^\text{k}+\mathbf{q}^\text{c}.
\eeq
The kinetic contributions $\mathsf{P}^\text{k}$ and $\mathbf{q}^\text{k}$ are given by
\beq
\label{4.14}
\mathsf{P}^\text{k}=\sum_{i=1}^s\int \dd\mathbf{v}\; m_i\mathbf{V}\mathbf{V}f_i(\mathbf{v}),
\eeq
\beq
\label{4.15}
\mathbf{q}^\text{k}=\sum_{i=1}^s\int \dd\mathbf{v}\; \frac{m_i}{2}V^2\mathbf{V}f_i(\mathbf{v}).
\eeq
The collisional transfer contributions are \cite{G19,GDH07}
\beqa
\label{4.16}
\mathsf{P}^\text{c}&=&\sum_{i,j=1}^s\sigma_{ij}^d m_{ij}\frac{1+\alpha_{ij}}{2}\int \dd\mathbf{v}_1\int \dd\mathbf{v}_2
\int \dd\widehat{\boldsymbol{\sigma}}\nonumber\\
& & \times \Theta\left(\widehat{\boldsymbol{\sigma}}\cdot\mathbf{g}_{12}\right)\left(\widehat{\boldsymbol{\sigma}}
\cdot\mathbf{g}_{12}\right)^2\widehat{\boldsymbol{\sigma}}\widehat{\boldsymbol{\sigma}}\int_{0}^{1}\dd x\nonumber\\
& & \times f_{ij}\Big(\mathbf{r}-x\boldsymbol{\sigma}_{ij},\mathbf{r}+(1-x)\boldsymbol{\sigma}_{ij},\mathbf{v}_1,\mathbf{v}_2,t\Big),
\eeqa
\beqa
\label{4.17}
\mathbf{q}^\text{c}&=&\sum_{i,j=1}^s\sigma_{ij}^dm_{ij}\frac{1+\alpha_{ij}}{8}\int \dd\mathbf{v}_1\int \dd\mathbf{v}_2\int \dd\widehat{\boldsymbol{\sigma}}\nonumber\\
& & \times \Theta\left(\widehat{\boldsymbol{\sigma}}\cdot\mathbf{g}_{12}\right)\left(\widehat{\boldsymbol{\sigma}}
\cdot\mathbf{g}_{12}\right)^2\widehat{\boldsymbol{\sigma}}\Big[4\left(\widehat{\boldsymbol{\sigma}}\cdot\mathbf{G}_{ij}\right)\nonumber\\
& & +
\left(\mu_{ji}-\mu_{ij}\right)\left(1-\alpha_{ij}\right)\left(\boldsymbol{\hat{\sigma}}\cdot\mathbf{g}_{12}\right)\Big]\int_{0}^{1}\dd x\nonumber\\
& &
\times f_{ij}\Big(\mathbf{r}-x\boldsymbol{\sigma}_{ij},\mathbf{r}+(1-x)\boldsymbol{\sigma}_{ij},\mathbf{v}_1,\mathbf{v}_2;t\Big).
\eeqa
Here, $m_{ij}=m_im_j/(m_i+m_j)$ is the reduced mass and $\mathbf{G}_{ij}=\mu_{ij}\mathbf{V}_1+\mu_{ji}\mathbf{V}_2$ is the velocity of the center of mass. Finally, the (total) cooling rate $\zeta$ due to inelastic collisions among all components is given by
\beqa
\label{2.25}
\zeta&=&\frac{1}{2dnT}\sum_{i,j=1}^s\sigma_{ij}^{d-1}m_{ij}\left(1-\alpha_{ij}^2\right)\int \dd\mathbf{v}_1\int \dd\mathbf{v}_2\int \dd\widehat{\boldsymbol{\sigma}}\nonumber\\
& & \times \Theta\left(\widehat{\boldsymbol{\sigma}}\cdot\mathbf{g}_{12}\right)\left(\widehat{\boldsymbol{\sigma}}
\cdot\mathbf{g}_{12}\right)^3f_{ij}
\left(\mathbf{r},\mathbf{r}+\boldsymbol{\sigma}_{ij},\mathbf{v}_1,\mathbf{v}_2;t\right).
\nonumber\\
\eeqa

As expected, the balance equations \eqref{4.9}--\eqref{4.11} are not a closed set of equations for the fields $n_i$, $\mathbf{U}$, and $T$. To transform these equations into a set of closed equations, one has to express the fluxes and the cooling rate in terms of the hydrodynamic fields and their gradients. The corresponding constitutive equations can be obtained by solving the set of Enkog kinetic equations \eqref{1} with the Chapman--Enskog method \cite{CC70} adapted to dissipative dynamics.

\section{First-order contributions to the partial temperatures}
\label{sec3}

The inelastic Enskog equation \eqref{1} was solved in Refs.\ \cite{GDH07,GHD07} by means of the Chapman--Enskog method. In particular, the first-order velocity distribution functions $f_i^{(1)}$ are given by \cite{GDH07}
\beqa
\label{5}
f_i^{(1)}&=&\boldsymbol{\mathcal{A}}_i\cdot\nabla\ln T+\sum_{j=1}^s\boldsymbol{\mathcal{B}}_{ij}\cdot \nabla\ln n_j \nonumber\\
& & + \mathcal{C}_{i,\lambda\beta}\frac{1}{2}\left(\partial_\lambda U_\beta+\partial_\beta U_\lambda-\frac{2}{d}\delta_{\lambda\beta}\nabla\cdot\mathbf{U}\right)\nonumber\\
& & +\mathcal{D}_i\nabla\cdot\mathbf{U},
\eeqa
where $\partial_\lambda\equiv \partial/\partial r_\lambda$. The unknowns $\boldsymbol{\mathcal{A}}_i(\mathbf{V})$, $\boldsymbol{\mathcal{B}}_{ij}(\mathbf{V})$, $\mathcal{C}_{i,\lambda\beta}(\mathbf{V})$, and $\mathcal{D}_i(\mathbf{V})$ are functions of the peculiar velocity $\mathbf{V}$ and they are the solutions of a set of coupled linear integral equations \cite{GDH07}. Approximate solutions to this set of integral equations were obtained in Refs. \cite{GHD07,MGH12} by considering the leading terms in a Sonine polynomial expansion. This procedure allows us to obtain explicit forms of the Navier--Stokes transport coefficients in terms of the mechanical parameters of the mixture (masses and sizes and the coefficients of restitution), the composition, and the density. Within the context of small gradients, the results apply in principle for arbitrary values of the coefficients of restitution and a wide range of densities.

However, as said in Sec.\ \ref{sec1}, the influence of the first-order contribution $T_i^{(1)}$ to $T_i$ on the transport coefficients was neglected in the above papers \cite{GDH07,GHD07}. This was essentially assumed because $T_i^{(1)}$ comes from the second-Sonine approximation and hence, it is expected that its impact on transport properties is small. Here, we want to determine $T_i^{(1)}$ to assess its influence on the bulk viscosity and the cooling rate.

According to Eq.\ \eqref{4.4}, the first-order contribution to the partial temperature $T_i$ is defined as
\beq
\label{6}
T_i^{(1)}=\frac{m_i}{d n_i}\int \dd\mathbf{v}\; V^2 f_i^{(1)}(\mathbf{V}).
\eeq
Since $T_i^{(1)}$ is a scalar, it can be only coupled to the divergence of the flow velocity $\nabla \cdot \mathbf{U}$ since $\nabla n$ and $\nabla T$ are vectors and $\partial_\lambda U_\beta+\partial_\beta U_\lambda-
(2/d)\delta_{\lambda\beta}\nabla\cdot\mathbf{U}$ is a traceless tensor. Thus, $T_i^{(1)}$ can be written as
$T_i^{(1)}=\varpi_i \nabla \cdot \mathbf{U}$, where
\beq
\label{8}
\varpi_i=\frac{m_i}{d n_i}\int\; \dd \mathbf{v}\;  V^2 \mathcal{D}_i(\mathbf{V}).
\eeq
The fact that the total temperature $T$ is not affected by the gradients implies necessarily the constraint
$\sum_{i=1}^s n_i T_i^{(1)}=0$. Thus, only $s-1$ partial temperatures are independent. The above constraint comes directly from the solubility condition
\beq
\label{9}
\sum_{i=1}^s \int \dd \mathbf{v} m_i V^2 f_i^{(1)}=0.
\eeq

As said before, apart from obtaining $T_i^{(1)}$, we are also interested here in revisiting previous calculations \cite{GDH07,GHD07} made for the bulk viscosity $\eta_\text{b}$ and the cooling rate $\zeta$. The first coefficient has only collisional contributions and its form can be identified by expanding the collisional transfer contribution $\mathsf{P}^\text{c}$ to the pressure tensor to first order in spatial gradients. A careful first-order expansion of the expression \eqref{4.16} to $\mathsf{P}^\text{c}$ gives the following form for $\eta_\text{b}$:
\beq
\label{9.1}
\eta_\text{b}=\eta_\text{b}'+\eta_\text{b}'',
\eeq
where
\beqa
\label{9.2}
\eta_\text{b}^{\prime}&=&\frac{\pi^{(d-1)/2}}{\Gamma\left(\frac{d+3}{2}\right)}\frac{d+1}{2d^{2}}\sum_{i=1}^{s}\sum_{j=1}^{s}m_{ij}
\left(1+\alpha_{ij}\right)\chi_{ij}^{(0)}\sigma_{ij}^{d+1}\nonumber\\
& & \times \int \dd\mathbf{v}_1\int \dd\mathbf{v}_{2}f_{i}^{(0)}(\mathbf{V}_{1})f_{j}^{(0)}(\mathbf{V}_{2})g_{12},
\eeqa
and
\beq
\label{9.3}
\eta_\text{b}''=-\frac{\pi^{d/2}}{d\Gamma\left(\frac{d}{2}\right)}\sum_{i=1}^{s}\sum_{j=1}^{s}\mu_{ji}
\left(1+\alpha_{ij}\right)\chi_{ij}^{(0)} n_i n_j\sigma_{ij}^{d}\varpi_i.
\eeq
In Eq.\ \eqref{9.2}, $f_i^{(0)}$ is the zeroth-order distribution. In addition, it is understood henceforth that the functional dependence of $\chi_{ij}^{(0)}(\mathbf{r},\mathbf{r}'|\{n_i\})$ on the compositions to zeroth order in the gradients has the same functional dependence on the densities replaced by $\{n_i\}\to \{n_i(\mathbf{r},t)\}$ at the point of interest.

The second contribution $\eta_\text{b}''$ to $\eta_\text{b}$ in Eq.\ \eqref{9.1} was neglected in previous works \cite{GDH07,GHD07,G19}. On the other hand, as said in Sec.\ \ref{sec1}, the contribution $\eta_\text{b}''$ was already accounted for in the studies on ordinary (elastic collisions) hard-sphere mixtures \cite{KS79a,KS79b,LCK83} carried out many years ago. In fact, for elastic collisions, Eq.\ \eqref{9.3} is consistent with Eq.\ (18a) of Ref.\ \cite{KS79b}.

In the case of the cooling rate, $\zeta\to \zeta^{(0)}+\zeta_U \nabla \cdot \mathbf{U}$ where
\beqa
\label{9.4}
& & \zeta^{(0)}=\frac{1}{2dnT}\sum_{i,j=1}^s\sigma_{ij}^{d-1}m_{ij}\left(1-\alpha_{ij}^2\right)\chi_{ij}^{(0)}\int \dd\mathbf{v}_1\int \dd\mathbf{v}_2\nonumber\\
& & \times \int \dd\widehat{\boldsymbol{\sigma}}\Theta\left(\widehat{\boldsymbol{\sigma}}\cdot\mathbf{g}_{12}\right)\left(\widehat{\boldsymbol{\sigma}}
\cdot\mathbf{g}_{12}\right)^3 f_i^{(0)}(\mathbf{r},\mathbf{v}_1,t)f_j^{(0)}(\mathbf{r},\mathbf{v}_2,t),
\nonumber\\
\eeqa
and $\zeta_U=\zeta^{(1,0)}+\zeta^{(1,1)}$. Here,
\beq
\label{15}
\zeta^{(1,0)}=-\frac{3\pi^{d/2}}{d^2\Gamma\left(\frac{d}{2}\right)}\sum_{i=1}^s\sum_{j=1}^s x_i n_j \mu_{ji}\sigma_{ij}^d \chi_{ij}^{(0)} \gamma_i(1-\al_{ij}^2),
\eeq
and the coefficient $\zeta^{(1,1)}$ is given in terms of the unknowns $\mathcal{D}_i$ as
\beqa
\label{16}
\zeta^{(1,1)}&=&\frac{1}{nT}\frac{\pi^{(d-1)/2}}{d\Gamma\left(\frac{d+3}{2}\right)}\sum_{i=1}^s\sum_{j=1}^s \sigma_{ij}^{d-1} \chi_{ij}^{(0)} m_{ij} (1-\al_{ij}^2)\nonumber\\
& & \times \int\dd \mathbf{v}_1\int\dd \mathbf{v}_2 \; g_{12}^3\; f_i^{(0)}(\mathbf{V}_1)\mathcal{D}_j (\mathbf{V}_2).
\eeqa
In Eq.\ \eqref{15}, $\gamma_i\equiv T_i^{(0)}/T$ is the temperature ratio of component $i$. The temperature ratios $\gamma_i$ verify the relation $\sum_i x_i \gamma_i=1$ and they are determined from the conditions $\zeta^{(0)}=\zeta_1^{(0)}=\zeta_2^{(0)}=\cdots =\zeta_s^{(0)}$,
where
\beq
\label{14.1}
\zeta_i^{(0)}=-\frac{m_i}{d n_i T_i^{(0)}}\sum_{j=1}^s\int \dd\mathbf{v} V^2 J_{ij}^{(0)}[f_i^{(0)},f_j^{(0)}].
\eeq

According to the results obtained in Ref.\ \cite{GDH07}, the coefficients $\varpi_i$ are the solutions of the set of coupled linear integral equations
\begin{widetext}
\beq
\label{10}
\frac{1}{2}\zeta^{(0)}\frac{\partial}{\partial {\bf V}}\cdot \left({\bf V}\mathcal{D}_{i}\right)+\frac{1}{2}\zeta^{(0)}\mathcal{D}_{i}
+\frac{1}{2}\zeta^{(1,1)}\frac{\partial}{\partial {\bf V}}\cdot \left({\bf V}f_{i}^{(0)}\right)-\sum_{j=1}^s\left(J_{ij}^{(0)}[\mathcal{D}_{i},f_j^{(0)}]+
J_{ij}^{(0)}[f_i^{(0)},\mathcal{D}_{j}]\right)=D_{i},
\eeq
where
\beq
\label{11}
J_{ij}^{(0)}[f_i^{(0)},f_j^{(0)}]=\chi_{ij}^{(0)} \sigma_{ij}^{d-1}\int \dd{\bf v}_{2}\int \dd\widehat{\boldsymbol{\sigma}}\,
\Theta (\widehat{{\boldsymbol {\sigma}}}\cdot {\bf g}_{12})(\widehat{\boldsymbol {\sigma }}\cdot {\bf g}_{12})\left[
\al_{ij}^{-2}f_i^{(0)}(\mathbf{v}_1'') f_j^{(0)}(\mathbf{v}_2'')
-f_i^{(0)}(\mathbf{v}_1) f_j^{(0)}(\mathbf{v}_2)\right]
\eeq
is the Boltzmann collision operator multiplied by the (constant) pair distribution function $\chi_{ij}^{(0)}$, and $D_i$ is given by
\beq
\label{12}
D_i\left(\mathbf{V}\right)=\frac{1}{2}\Bigg[\frac{2}{d}\left(1-p^*\right)-\zeta^{(1,0)}\Bigg]
\frac{\partial}{\partial \mathbf{V}}\cdot \left(\mathbf{V}f_i^{(0)}\right)-f_i^{(0)}+\sum_{j=1}^s \Bigg(n_j \frac{\partial f_i^{(0)}}
{\partial n_j} +
\frac{1}{d}\mathcal{K}_{ij,\beta}\left[\frac{\partial f_i^{(0)}}{\partial V_\beta}\right]\Bigg).
\eeq
In addition, $\boldsymbol{\mathcal{K}}_{ij}[X_j]$ is the collision operator
\beq
\label{13}
\boldsymbol{\mathcal{K}}_{ij}[X_j] =\sigma_{ij}^{d}\chi_{ij}^{(0)} \int \dd \mathbf{v}_{2}\int \dd\widehat{\boldsymbol {\sigma
}}\Theta (\widehat{\boldsymbol {\sigma}} \cdot
\mathbf{g}_{12})(\widehat{\boldsymbol {\sigma }}\cdot
\mathbf{g}_{12})\widehat{\boldsymbol {\sigma}}\left[ \alpha_{ij}
^{-2}f_i^{(0)}(\mathbf{v}_{1}'')X_j(\mathbf{v}_{2}'')+f_i^{(0)}(\mathbf{v}_{1})X_j(\mathbf{v}_{2})\right],
\eeq
\end{widetext}
and the (reduced) hydrostatic pressure $p^*\equiv p/(n T)$ is
\beq
\label{14}
p^*=1+\frac{\pi^{d/2}}{d\Gamma\left(\frac{d}{2}\right)}\sum_{i=1}^s\sum_{j=1}^s\; \mu_{ji}x_i x_j n \sigma_{ij}^d \chi_{ij}^{(0)} \gamma_i (1+\al_{ij}).
\eeq

Since $\mathcal{D}_i(\mathbf{V}) \propto D_i(\mathbf{V})$, the solubility condition \eqref{9} requires necessarily that
\beq
\label{17}
\sum_{i=1}^s \int \dd \mathbf{v} m_i V^2 D_i(\mathbf{V})=0.
\eeq
This condition can easily be verified by direct integration of Eq.\ \eqref{12} and using Eqs.\  \eqref{15}--\eqref{14}, the relation $\sum_i x_i \gamma_i=1$, and the result
\beqa
\label{18}
A_{i}&\equiv& \sum_{j=1}^s\int \dd\mathbf{v} m_i V^2 \mathcal{K}_{ij,\lambda}\left[\frac{\partial f_j^{(0)}}{\partial V_\lambda}\right]\nonumber\\
&=&-
\frac{\pi^{d/2}}{\Gamma\left(\frac{d}{2}\right)}T\sum_{j=1}^s\chi_{ij}^{(0)} n_i n_j \sigma_{ij}^d (1+\al_{ij})\Bigg[3\mu_{ji}(1+\al_{ij})\nonumber\\
& & \times \left(\frac{\gamma_i}{m_i}+\frac{\gamma_j}{m_j}\right)-4\frac{\gamma_i}{m_i}\Bigg].
\eeqa

In the low-density regime ($n_i\sigma_{ij}^d\to 0$), $p^*=1$, $\zeta^{(1,0)}=0$, the combination $\sum_j n_j \partial f_i^{(0)}/\partial n_j-f_i^{(0)}$ and the operator $\boldsymbol{\mathcal{K}}_{ij}[X_j]$ vanish, and so $D_i=0$ in the integral equation \eqref{10}. This means $\mathcal{D}_i=0$ and hence, the first-order contributions $\varpi_i$ to the partial temperatures vanish for \emph{dilute} granular mixtures. This agrees with the previous results obtained in the low-density regime \cite{GD02,SGNT06,GM07,SNTG09}.

\section{Leading Sonine approximation}
\label{sec4}

It is quite apparent that the calculation of $\varpi_i$ requires to solve the integral equation \eqref{10} as well as to know the zeroth-order distributions $f_i^{(0)}$. With respect to the latter, previous results \cite{GD99b,MG02} derived for homogeneous states have clearly shown that, in the region of thermal velocities, $f_i^{(0)}$ is well represented by the Maxwellian velocity distribution defined at the lowest-order partial temperature $T_i^{(0)}$, namely,
\beq
\label{19}
f_i^{(0)}(\mathbf{V})\to f_{i,\text{M}}(\mathbf{V})=n_i \left(\frac{m_i}{2\pi T_i^{(0)}}\right)^{d/2}\exp\left(-\frac{m_i V^2}{2T_i^{(0)}}\right).
\eeq
This means that we neglect here non-Gaussian corrections to the distributions $f_i^{(0)}$ and hence, one expects to get simple but accurate expressions for the transport coefficients. With this approximation, $\zeta_i^{(0)}$ is
\beqa
\label{20}
\zeta_{i}^{(0)}&=&\frac{4\pi^{\left(d-1\right)/2}}{d\Gamma\left(\frac{d}{2}\right)}v_0
\sum_{j=1}^s\;n_{j}\chi_{ij}^{(0)}\mu_{ji}\sigma_{ij}^{d-1}\left(1+\alpha_{ij}\right)\nonumber\\
& & \times \left(\frac{\beta_i+\beta_j}
{\beta_i\beta_j}\right)^{1/2} \Big[1-\frac{\mu_{ji}}{2}\left(1+\alpha_{ij}\right)\frac{\beta_i+\beta_j}{\beta_j}\Big],
\nonumber\\
\eeqa
where $v_0(T)=\sqrt{2T/\overline{m}}$ is a thermal speed of the mixture, $\overline{m}=\sum_i m_i/s$, and $\beta_i=m_i T/\overline{m} T_i^{(0)}$. Furthermore, according to Eq.\ \eqref{9.2} the contribution $\eta_\text{b}'$ to the bulk viscosity can also be computed by using the Maxwellian approximation \eqref{19} with the result
\beqa
\label{20.1}
\eta_\text{b}^{\prime}&=&\frac{\pi^{(d-1)/2}}{d^2\Gamma\left(\frac{d}{2}\right)}v_0\sum_{i=1}^{2}\sum_{j=1}^{2}m_{ij}
\left(1+\alpha_{ij}\right)\chi_{ij}^{(0)} n_i n_j \sigma_{ij}^{d+1}\nonumber\\
& & \times \left(\frac{\beta_i+\beta_j}{\beta_i\beta_j}\right)^{1/2}.
\eeqa

To solve the integral equation \eqref{10}, one takes the leading Sonine approximation to $\mathcal{D}_i(\mathbf{V})$
\beq
\label{21}
\mathcal{D}_i(\mathbf{V})\rightarrow f_{i\text{M}}(\mathbf{V})W_i(\mathbf{V})\frac{\varpi_i}{T_i^{(0)}},
\eeq
where
\beq
\label{21.1}
W_i(\mathbf{V})=\frac{m_iV^2}{2T_i^{(0)}}-\frac{d}{2}.
\eeq
The relation between $\zeta^{(1,1)}$ and $\varpi_i$ can be easily obtained by substitution of Eq.\ \eqref{21} into Eq.\ \eqref{16}. The result is
\beq
\label{22}
\zeta^{(1,1)}=\sum_{i=1}^s\; \xi_i \varpi_i,
\eeq
where
\beqa
\label{23}
\xi_i&=&\frac{3\pi^{(d-1)/2}}{2d\Gamma\left(\frac{d}{2}\right)}\frac{v_0^3}{n T T_i^{(0)}}\sum_{j=1}^s n_i n_j \sigma_{ij}^{d-1}
\chi_{ij}^{(0)} m_{ij}(1-\al_{ij}^2) \nonumber\\
& & \times \left(\beta_i+\beta_j\right)^{1/2}\beta_i^{-3/2}\beta_j^{-1/2}.
\eeqa
The coefficients $\varpi_i$ can be finally obtained by substituting Eq.\ \eqref{21} into Eq.\ \eqref{10}, multiplying it with $m_iV^2$ and integrating over the velocity. After some algebra, the corresponding set of coupled linear algebraic equations for the coefficients $\varpi_i$ are given by
\beq
\label{24}
\sum_{j=1}^s\Big(\omega_{ij}+\frac{1}{2}\zeta^{(0)}\delta_{ij}+T_i^{(0)}\xi_j\Big)\varpi_j
=B_i,
\eeq
where
\beq
\label{24.1}
B_i=\frac{2}{d}T_i^{(0)}\left(1-p^*\right)-T_i^{(0)}\zeta^{(1,0)}-T\phi \frac{\partial \gamma_i}{\partial \phi}-\frac{A_i}{d^2 n_i},
\eeq
and 
\beq
\label{24.2} 
\phi=\frac{\pi^{d/2}}{2^{d-1}d\Gamma \left(\frac{d}{2}\right)} \sum_{i=1}^s\; n_i\sigma_i^d
\eeq
is the solid volume fraction. Upon obtaining Eq.\ \eqref{24.1} we have taken into account that the dependence of the temperature ratios $\gamma_i$ on the densities $n_i$ is through their dependence on the mole fractions $x_i$ and the volume fraction $\phi$. Furthermore, the collision frequencies $\omega_{ij}$ are defined as
\beqa
\label{25}
\omega_{ii}&=&\frac{1}{dn_iT_i^{(0)}}\Bigg(\sum_{j=1}^s\int\dd\mathbf{v}m_iV^2J_{ij}^{(0)}\left[f_{i,\text{M}}W_i,f_j^{(0)}\right]\nonumber\\
& &
+\int\dd\mathbf{v}m_iV^2J_{ii}^{(0)}\left[f_i^{(0)},f_{i,\text{M}}W_i\right]\Bigg) ,
\eeqa
\beq
\label{26}
\omega_{ij}=\frac{1}{dn_iT_j^{(0)}}\int\dd\mathbf{v}m_iV^2J_{ij}^{(0)}\left[f_i^{(0)},f_{j,\text{M}}W_j\right], \quad (i\neq j).
\eeq
In the Maxwellian approximation \eqref{19}, $\omega_{ii}$ and $\omega_{ij}$ are
\beqa
\label{27}
\omega_{ii}&=&-\frac{\pi^{(d-1)/2}}{2dT_i^{(0)}\Gamma\left(\frac{d}{2}\right)}v_0^3\Bigg\{\frac{3}{\sqrt{2}}
n_i\sigma_i^{d-1}m_i \chi_{ii}^{(0)} \beta_i^{-3/2}
\left(1-\alpha_{ii}^2\right)\nonumber\\
& & -\sum_{j\neq i}^s n_j m_{ij} \sigma_{ij}^{d-1}\chi_{ij}^{(0)} \left(1+\alpha_{ij}\right)\left(\beta_i+\beta_j\right)^{-1/2}\beta_i^{-3/2}\nonumber\\
& & \times \beta_j^{-1/2}
\Big[3\mu_{ji}\left(1+\alpha_{ij}\right)\left(\beta_i+\beta_j\right)-2\left(2\beta_i+3\beta_j\right)\Big]\Bigg\},
\nonumber\\
\eeqa
\beqa
\label{28}
\omega_{ij}&=&\frac{\pi^{(d-1)/2}}{2dT_j^{(0)}\Gamma\left(\frac{d}{2}\right)}v_0^3n_jm_{ij}\sigma_{ij}^{d-1}\chi_{ij}^{(0)}
\left(1+\alpha_{ij}\right)\nonumber\\
& & \times \left(\beta_i+\beta_j\right)^{-1/2}
\beta_i^{-1/2}\beta_j^{-3/2}\Big[3\mu_{ji}\left(1+\alpha_{ij}\right)\nonumber\\
& & \times
\left(\beta_i+\beta_j\right)-2\beta_j\Big].
\eeqa
In Eqs.\ \eqref{27}--\eqref{28}, it is understood that $i\neq j$. The set of algebraic equations \eqref{25} can be now easily solved. In particular, for a binary mixture ($s=2$) the solution of Eq.\ \eqref{24} for $\varpi_1$ can be written as  
\beq
\label{29}
\varpi_1=\frac{B_1}{\omega_{11}-\frac{x_1}{x_2}\omega_{12}+\frac{1}{2}\zeta^{(0)}+T_1^{(0)}\left(\xi_1-\frac{x_1}{x_2}\xi_2\right)},
\eeq
where the relation $\varpi_2=-(x_1/x_2) \varpi_1$ has been accounted for. The expression for $\varpi_2$ can be easily obtained from Eq.\ \eqref{29} by making the changes $1\leftrightarrow 2$. The solution \eqref{29} is indeed consistent with the requirement $x_1 \varpi_1+x_2 \varpi_2=0$. This is because $x_1\gamma_1+x_2\gamma_2=1$, $B_2=-(x_1/x_2)B_1$, and $\omega_{11}-(x_1/x_2)\omega_{12}+\xi_1/x_1=\omega_{22}-(x_2/x_1)\omega_{21}+\xi_2/x_2$.

The expression \eqref{29} provides $\varpi_1$ in terms of the parameters of the mixture. Its explicit form is relatively long and is omitted here for the sake of brevity. A simple but interesting case corresponds to ordinary mixtures (elastic collisions) where $\zeta^{(0)}=0$, $\xi_i=0$, $\gamma_i=1$, $\beta_1=2\mu_{12}$, $\beta_2=2\mu_{21}$, and $\varpi_1$ is
\beqa
\label{30}
\varpi_1&=&\frac{4\pi^{d/2}}{d^2\Gamma\left(\frac{d}{2}\right)}T \left(\omega_{11}-\frac{x_1}{x_2}\omega_{12}\right)^{-1}\Bigg[n_2\sigma_{12}^d\chi_{12}^{(0)}\big(x_2\mu_{21}\nonumber\\
& & -x_1\mu_{12}\big)
+\frac{1}{2}x_2\left(n_1\sigma_1^d \chi_{11}^{(0)}-n_2\sigma_2^d \chi_{22}^{(0)}\right)\Bigg].
\eeqa
Equation \eqref{30} differs from the one obtained by Jenkins and Mancini \cite{JM87} for nearly elastic hard spheres ($d=3$). This discrepancy is essentially due to the fact that the distribution functions of each species in Ref.\ \cite{JM87} are assumed to be Maxwellian distributions even in inhomogeneous situations. This was already noted by the authors of this paper since they conclude that their expression for $\varpi_1$ could be improved by determining the perturbations to the Maxwellians using the Chapman--Enskog procedure \cite{CC70}. Expression \eqref{30} accounts for not only the different centers $\mathbf{r}$ and $\mathbf{r}\pm \boldsymbol{\sigma}_{ij}$ of the colliding pair in the Enskog collision operator \eqref{2} but also for the form of the first-order distribution $f^{(1)}$ given by Eq.\ \eqref{5}.

On the other hand, for a three-dimensional system ($d=3$), the expression \eqref{30} for $\varpi_1$ agrees with the one derived in Ref.\ \cite{KS79b} (see Eq.\ (22d) of \cite{KS79b}) for a hard-sphere binary mixture. This confirms the relevant known limiting cases for the granular mixture results derived here for the temperature ratios.

Once the first-order contributions to the partial temperatures are known, the first-order contribution $\zeta_U$ to the cooling rate can be explicitly obtained by employing Eqs.\ \eqref{15}, \eqref{16}, and \eqref{22}--\eqref{23}. In addition, the second contribution $\eta_\text{b}''$ to the bulk viscosity $\eta_\text{b}$ can be obtained from Eq.\ \eqref{9.3}. Thus, $\eta_\text{b}=\eta_\text{b}'+\eta_\text{b}''$ is completely determined from Eqs.\ \eqref{9.3} and \eqref{20.1}. For elastic collisions ($\al_{ij}=1$), as expected the corresponding expression for $\eta_\text{b}$  is consistent with previous works on ordinary mixtures \cite{KS79a,KS79b,LCK83}.

\section{Binary granular mixtures}
\label{sec5}

In order to illustrate the dependence of the coefficients $\varpi_i$, $\zeta_U$, and $\eta_\text{b}$ on the parameter space of the system, a binary  mixture ($s=2$ and so, $\varpi_2=-x_1 \varpi_1/x_2$) of inelastic hard spheres ($d=3$) is considered. The above coefficients depend on many parameters: $\left\{x_1,T, m_1/m_2, \sigma_1/\sigma_2, \phi, \al_{11}, \al_{22}, \al_{12} \right\}$. A similar complexity also exists in the elastic limit \cite{LCK83}, so the relevant new feature is the dependence of $\varpi_1$, $\zeta_U$, and $\eta_\text{b}$ on the coefficients of restitution. Moreover, for the sake of simplicity, the case of a common coefficient of restitution ($\al_{11}=\al_{22}=\al_{12}\equiv \al$) of an equimolar mixture ($x_1=\frac{1}{2}$) with $\sigma_1=\sigma_2$ and solid volume fraction $\phi=0.2$ (moderately dense gas) is considered. This reduces the parameter space to three quantities: $\left\{T, m_1/m_2,\al\right\}$. The dependence on temperature can be scaled out by introducing the (dimensionless) quantities $\varpi_1^*=(n \sigma_{12}^2 v_0/T) \varpi_1$ and $\eta_\text{b}^*\equiv \eta_\text{b}(\alpha)/\eta_\text{b}(1)$, where $\eta_\text{b}(1)$ is the bulk viscosity for elastic collisions. The coefficient $\zeta_U$ is dimensionless.

\begin{figure}
\includegraphics[width=0.9\columnwidth]{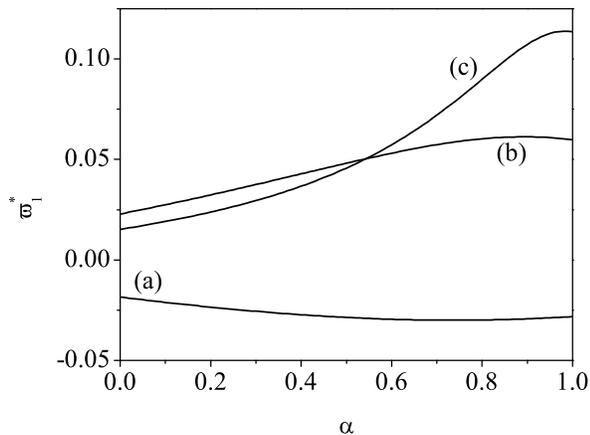}
\caption{Plot of the (reduced) coefficient $\varpi_1^*$ as a function of the common coefficient of restitution $\alpha$ for a binary mixture of hard spheres ($d=3$) with $x_1=\frac{1}{2}$, $\sigma_1=\sigma_2$, $\phi=0.2$, and three different values of the mass ratio $m_1/m_2$: $m_1/m_2=0.5$ (a), $m_1/m_2=4$ (b), and $m_1/m_2=10$ (c).}
\label{fig1}
\end{figure}
\begin{figure}
\includegraphics[width=0.9\columnwidth]{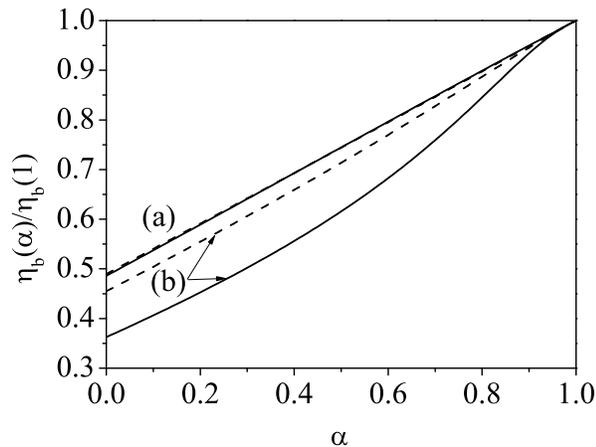}
\caption{Plot of the (reduced) bulk viscosity $\eta_\text{b}(\alpha)/\eta_\text{b}(1)$ as a function of the common coefficient of restitution $\alpha$ for a binary mixture of hard spheres ($d=3$) with $x_1=\frac{1}{2}$, $\sigma_1=\sigma_2$, $\phi=0.2$, and two different values of the mass ratio $m_1/m_2$: $m_1/m_2=0.5$ (a) and $m_1/m_2=10$ (b). The dashed lines are the results for the (reduced) bulk viscosity when the contribution $\eta_\text{b}''$ to $\eta_\text{b}$ is neglected.}
\label{fig2}
\end{figure}
\begin{figure}
\includegraphics[width=0.9\columnwidth]{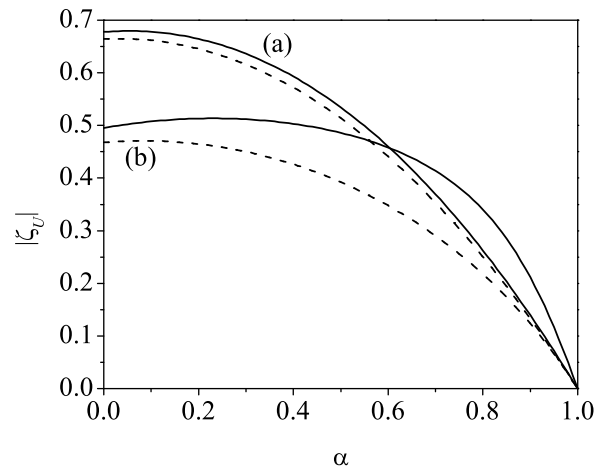}
\caption{Plot of the magnitude of the (reduced) coefficient $\zeta_U$ as a function of the common coefficient of restitution $\alpha$ for a binary mixture of hard spheres ($d=3$) with $x_1=\frac{1}{2}$, $\sigma_1=\sigma_2$, $\phi=0.2$, and two different values of the mass ratio $m_1/m_2$: $m_1/m_2=0.5$ (a) and $m_1/m_2=10$ (b). The dashed lines are the results for the coefficient $\zeta_U$ when the contribution $\zeta^{(1,1)}$ to $\zeta_U$ is neglected.}
\label{fig3}
\end{figure}

To display the dependence of the coefficients $\varpi_1^*$, $\eta_\text{b}^*$, and $\zeta_U$ on $\alpha$, we have still to provide the form for the pair distribution function $\chi_{ij}^{(0)}$. In the case of spheres ($d=3$), a good approximation of $\chi_{ij}^{(0)}$ is \cite{B70,GH72,LL73}
\beq
\label{35}
\chi_{ij}^{(0)}=\frac{1}{1-\phi}+\frac{3}{2}\frac{\phi}{(1-\phi)^2}\frac{\sigma_i \sigma_j M_2}{\sigma_{ij}M_3}+\frac{1}{2}\frac{\phi^2}{(1-\phi)^3}
\left(\frac{\sigma_i \sigma_j M_2}{\sigma_{ij}M_3}\right)^2,
\eeq
where $M_\ell=\sum_i x_i \sigma_i^\ell$. In Fig.\ \ref{fig1}, the (reduced) coefficient $\varpi_1^*$ is plotted as a function of the coefficient of restitution $\al$ for several values of the mass ratio. It is quite apparent first that the influence of the inelasticity on $\varpi_1^*$ is significant, specially for high mass ratios. With respect to the dependence on the mass ratio, we see that while $\varpi_1^*$ increases with inelasticity when $m_1/m_2<1$, the opposite happens when $m_1/m_2>1$. Furthermore, Fig.\ \ref{fig1} also highlights that the magnitude of the first-contribution to the partial temperature is in general quite small in comparison with the values of the remaining transport coefficients of the mixture \cite{GHD07,MGH12}. To assess the impact of $\varpi_1^*$ on the bulk viscosity and the first-order contribution to the cooling rate, Figs.\ \ref{fig2} and \ref{fig3} show the $\al$-dependence of the (reduced) coefficients $\eta_\text{b}(\alpha)/\eta_\text{b}(1)$ and $\zeta_U$, respectively, for two values of the mass ratio. We also plot the corresponding values of these coefficients when $\varpi_1^*$ is neglected. Although both predictions (with and without $\varpi_1^*$) agree qualitatively, we observe that the effect of $\varpi_1^*$ on both transport coefficients cannot be neglected specially for high mass ratios and moderate inelasticity (let's say, for instance $\al \simeq 0.6$). This means that previous results \cite{GDH07,GHD07} derived for both the bulk viscosity and the cooling rate of granular mixtures must be slightly changed when the masses of the constituents of the mixture become very disparate and/or the collisional dissipation becomes significant.

\section{Concluding remarks}
\label{sec6}

One of the most intriguing differences between ordinary and granular mixtures is the absence of energy equipartition in homogeneous states.
This means that the zeroth-order contributions $T_i^{(0)}$ to the \emph{partial} temperatures $T_i$ (measuring the mean kinetic energy of each species) of granular mixtures are different for mechanically different components, reflecting a violation of the equipartition theorem valid for elastic collisions \cite{GD99b}. The origin of this violation is the inelasticity in collisions and its impact on transport problems such as thermal diffusion segregation \cite{G08a,G09,G11} has been shown to be quite significant, specially for strong dissipation and/or disparate mass ratios.

In addition, as was already noted in some of the pioneering papers of the Enskog theory for multicomponent ordinary mixtures \cite{KS79a,KS79b,LCK83}, a breakdown of energy equipartition is also present in the Navier--Stokes domain (first-order in spatial gradients) for moderately dense mixtures. The origin of this violation is associated with the spatial gradients, and more specifically with the divergence of flow velocity since the first-order contributions $T_i^{(1)}$ to the partial temperatures are proportional to $\nabla\cdot \mathbf{U}$. This additional source of energy nonequipartition is independent of the one appearing in the homogeneous cooling state for granular mixtures.

On the other hand, the coefficients $T_i^{(1)}$ are usually neglected in many of the works devoted to granular mixtures \cite{JM89,GDH07,GHD07} because only the first terms in the Sonine polynomial expansion are retained. Since $T_i^{(1)}\propto \nabla \cdot \mathbf{U}$, an interesting question is to assess the impact of the first-order coefficients $T_i^{(1)}$ on both the bulk viscosity $\eta_\text{b}$ and the first-order contribution $\zeta_U$ to the cooling rate.

The goal of this paper has been to determine the coefficients $T_i^{(1)}$ from the Chapman--Enskog solution to the (inelastic) version of the Enskog kinetic equation \cite{G19}. As in previous works \cite{GDH07,GHD07}, this task has been achieved in two different steps. First, we have obtained in an \emph{exact} way the set of linear integral equations that the first-order contributions $T_i^{(1)}$ satisfy. This has allowed to prove the solubility condition for solving this set of integral equations. As a second step, an approximate solution to the above set of equations is required for practical purposes in order to explicitly express the coefficients $T_i^{(1)}$ in terms of the parameter space of the problem (masses, diameters, composition, density, and coefficients of restitution). This task has been achieved by considering the leading terms in the Sonine polynomial expansion. Thus, the results derived here for $T_i^{(1)}$ extend to inelastic collisions the calculations performed many years ago \cite{KS79a,KS79b,LCK83} for ordinary hard-sphere mixtures. Moreover, the expressions obtained here for $\eta_\text{b}$ [given by Eqs.\ \eqref{9.1}--\eqref{9.3}] and $\zeta_U$ [given by Eqs.\ \eqref{15}, \eqref{22}, and \eqref{23}] correct the previous results derived in Refs.\ \cite{GDH07,GHD07} where the contributions $\eta_\text{b}''$ and $\zeta^{(1,1)}$ to $\eta_\text{b}$ and $\zeta_U$, respectively, were implicitly neglected.

For the sake of illustration and to assess the impact of $T_i^{(1)}$ on $\eta_\text{b}$ and $\zeta_U$, a binary mixture with a common coefficient of restitution ($\al_{ij}\equiv \al$) has been considered to analyze the dependence of the above transport coefficients on inelasticity. First, as Fig.\ \ref{fig1} shows, we observe that the effect of inelasticity on the first-order contributions to the partial temperatures is in general quite important, specially for large mass ratios. With respect to the influence of $T_1^{(1)}$ on  $\eta_\text{b}$ and $\zeta_U$, Figs.\ \ref{fig2} and \ref{fig3} highlight that the impact of the first-order partial temperature on both the bulk viscosity and the cooling rate can be relatively important for moderate inelasticity and/or disparate mass ratios.

An interesting problem is to extend the present results to the case of polydisperse granular mixtures driven by a stochastic bath with friction \cite{PLMPV98,SVGP10,SVCP10}. This kind of thermostats models the effect of the surrounding interstitial viscous gas on the dynamics of grains (granular suspensions). An extensive study on the transport coefficients for driven granular mixtures at low density has been carried out in Refs.\ \cite{KG13,KG18,KG19}. In contrast with the findings reported here for \emph{freely} cooling granular dilute gases (where $T_i^{(1)}=0$ when $\phi=0$), the results derived for driven systems \cite{KG19} show that the first-order contributions $T_i^{(1)}$ to the partial temperatures are different from zero even when $\phi=0$. The extension of the results obtained in Refs.\ \cite{KG13,KG18,KG19} to finite density is an interesting project. Work along this line will be worked out in the near future.

In summary, we have revisited previous works on polydisperse granular mixtures \cite{GDH07,GHD07} where the first-order contributions $T_i^{(1)}$ to the partial temperatures were neglected. The present work fixes the above limitation by including not only the calculation of $T_i^{(1)}$ but also their influence on the bulk viscosity $\eta_\text{b}$ and on the first-order contribution $\zeta_U$ to the cooling rate. Our results show first that the first-order coefficients $T_i^{(1)}$  exhibit in general a complex dependence on the coefficients of restitution of the mixture. In addition, they also show that the impact of $T_i^{(1)}$ on both $\eta_\text{b}$ and $\zeta_U$ cannot be neglected for disparate masses and/or strong dissipation. In this context, the results derived before for polydisperse dense granular mixtures \cite{GDH07,GHD07} must be slightly modified by including the contributions coming from the partial temperatures $T_i^{(1)}$ to the transport properties and the cooling rate.

\acknowledgments

We are grateful to Dr.\ Mariano L\'opez de Haro for a critical reading of the manuscript and for calling our attention to the papers \cite{KS79a} and \cite{KS79b}. The present work has been supported by the Spanish Government through Grant No. FIS2016-76359-P and by the Junta de Extremadura (Spain) Grant Nos. IB16013 (V.G.) and GR18079, partially financed by ``Fondo Europeo de Desarrollo Regional'' funds. The research of Rub\'en G\'omez Gonz\'alez has been supported by the predoctoral fellowship BES-2017-079725 from the Spanish Government.

\end{document}